# US Velocimetry in Participants with Aortoiliac Occlusive Disease

Stefan Engelhard
Majorie van Helvert
Jason Voorneveld
Johan G Bosch
Guillaume PR Lajoinie
Michel Versluis
Erik Groot Jebbink
Michel MPJ Reijnen

# Abstract

**Background:** The accurate quantification of blood flow in aortoiliac arteries is challenging but clinically relevant because local flow patterns can influence atherosclerotic disease.

**Purpose:** To investigate the feasibility and clinical application of two-dimensional blood flow quantification using high-frame-rate contrast-enhanced US (HFR-CEUS) and particle image velocimetry (PIV), or US velocimetry, in participants with aortoiliac stenosis.

**Materials and Methods:** In this prospective study, participants with a recently diagnosed aortoiliac stenosis underwent HFR-CEUS measurements of the pre- and poststenotic vessel segments (August 2018 to July 2019). Two-dimensional quantification of blood flow was achieved by performing PIV analysis, which was based on pairwise cross-correlation of the HFR-CEUS images. Visual inspection of the entire data set was performed by five observers to evaluate the ability of the technique to enable adequate visualization of blood flow. The contrast-to-background ratio and average vector correlation were calculated. In two participants who showed flow disturbances, the flow complexity and vorticity were calculated.

**Results:** 35 participants (median age, 67 years; age range, 56–84 years; 22 men) were included. Visual scoring showed that flow quantification was achieved in 41 of 42 locations. In 25 locations, one or multiple issues occurred that limited optimal flow quantification, including loss of correlation during systole (n = 12), shadow regions (n = 8), a short vessel segment in the image plane (n = 7), and loss of contrast during diastole (n = 5). In the remaining 16 locations, optimal quantification was achieved. The contrast-to-background ratio was higher during systole than during diastole (11.0 6 2.9 vs 6.9 6 3.4, respectively; $p < 0.001$), whereas the vector correlation was lower (0.58 6 0.21 vs 0.47 6 0.13; $p < 0.001$). Flow complexity and vorticity were high in regions with disturbed flow.

**Conclusion:** Blood flow quantification with US velocimetry is feasible in patients with an aortoiliac stenosis, but several challenges must be overcome before implementation into clinical practice.

# Introduction

Time-resolved quantification of blood flow in diseased aortoiliac regions is challenging because of complex flow patterns near the aortic bifurcation and around stenosis. These flow patterns could be used to improve the assessment of stenosis severity and predict disease progression. For example, blood flow patterns have been correlated with the development and progression of atherosclerotic plaques. Specifically, lesions are more likely to form in areas of low wall shear stress [1], [2], which induces major changes in endothelial cells, making the vessel wall more prone to atherosclerosis [3], [4].

The peak systolic velocity ratio obtained with duplex US is traditionally applied to quantify the severity of a stenosis. This parameter shows mixed results compared with the reference standard, the invasively measured pressure gradient over the stenosis [5], [6]. This discrepancy can be explained by the angle-dependency of DUS, only providing a one-dimensional blood flow velocity estimate along the transducer axis, in a complex anatomic region where assumptions about flow direction are often inaccurate [7], [8].

High-frame-rate contrast-enhanced US (HFR-CEUS) combined with particle image velocimetry (PIV), or US velocimetry (echoPIV), enables two-dimensional angle-independent blood flow quantification. EchoPIV could be used to improve and expand the evaluation of lesion severity and to predict atherosclerotic disease progression. A previous study showed that blood flow quantification in the aortoiliac region with echoPIV is feasible in healthy volunteers [9]. However, US imaging in patients with atherosclerosis is more challenging, due to elongated and calcified arteries. This study aimed to investigate the feasibility and clinical application of two-dimensional blood flow quantification using echoPIV, in participants with aortoiliac stenosis.

## Materials and Methods

### Study Design

This prospective study was conducted in accordance with Good Clinical Practice guidelines, approved by an institutional review board (NL63077.091.17), and registered with the Netherlands Trial Register (NTR6980). Thirty-five consecutive participants aged over 50 years with intermittent claudication (Rutherford category of 1–3) based on aortoiliac stenosis were included after providing written informed con-sent. Participants were excluded from the study if the use of contrast microbubbles was contraindicated. Demographic and clinical data were retrieved from electronic health records [10], [11]. HFR-CEUS was performed between August 2018 and July 2019 within 1 month after diagnosis. Contrast-enhanced CT scans (section thickness, 0.4 mm) were obtained as an anatomic reference. Agatston calcium scores were calculated by performing automatic calcium segmentation (Intuition, TeraRecon), with a threshold of 600 HU being used to account for the contrast agent-filled vessel lumen [12].

## HFR-CEUS Measurements

HFR-CEUS was performed with a Vantage 256 Research US System (Verasonics, Kirkland, WA), and a curved array transducer (GE C1-6D, General Electric Company, Boston, MA). Prior to HFR-CEUS, blood flow velocities were measured with DUS, using an iU22 US machine (Philips Healthcare, Best, the Netherlands).

At each location, two doses of contrast microbubbles (SonoVue; Bracco, Milan, Italy) were administered (9). Microbubble arrival was monitored by using the Verasonics system, with a live imaging sequence at 100 frames/sec. When a quasi-stable concentration of contrast agent was visually established, two HFR-CEUS measurements were performed with a mechanical index of 0.05 and 0.1, both with a center frequency of 2.2MHz and a pulse length of one cycle. mages were captured for 2.5 seconds at 2000 frames/sec with use of a three-angled diverging wave acquisition scheme (pulse repetition frequency, 6000 Hz). Subsequent injections were given after complete washout of the contrast agent on the live images (2–10 minutes after injection). Four measurements were obtained for each location (0.5-mL and 1-mL contrast, each acquired with mechanical indexes of 0.05 and 0.1). This protocol was derived from a previous study [9].

## Data Analysis

Data were processed off-line with Matlab (R2019b, MathWorks). Raw HFR-CEUS data were reconstructed into images by using coherent compounding of the three transmit angles, increasing contrast and resolution [13]. Clutter suppression was performed using a singular value decomposition based filter [14]. Rank selection (ie the cut-off between tissue, blood, and noise) was performed automatically [15]. PIV analysis was performed by using a custom implementation in Matlab consisting of two iterations with a square block size of 5.6 mm and two iterations of 2.8 mm with 75% overlap, resulting in a 0.69-mm vector resolution. To improve the signal-to-noise ratio, correlation averaging was performed over 20 frames, resulting in 100 velocity fields per second. All velocity data and a selection of HFR-CEUS data were stored in a repository that can be accessed (with author permission) at *https://www.doi.org/10.4121/c.5497704*.

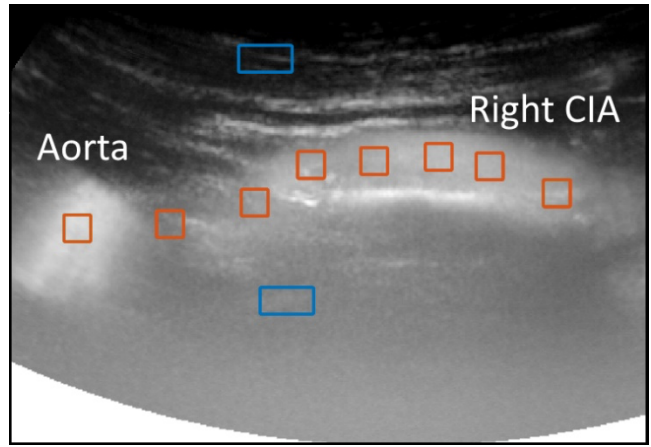

**Figure 1: Average of 100 high-frame-rate contrast-enhanced US images after postprocessing. To calculate the contrast-to-background ratio, eight regions are selected inside the vessel (orange squares), and one region is selected above and below the vessel as the background (blue squares). CIA = common iliac artery.**

## Feasibility Scoring

Qualitative scoring of all velocity data was performed by means of visual inspection by five investigators with expertise in blood flow imaging research (M.M.P.J.R., vascular surgeon with 18 years of experience; E.G.J., assistant professor; J.V., postdoctoral researcher; and S.E. and M.V., PhD candidates). For each location, the best measurement was selected and used to assess feasibility, which was classified according to three categories: un-feasible (no meaningful information could be obtained), partial quantification (blood flow was adequately visualized, but this only occurred during part of the heart cycle or in a subregion of the imaged vessel because one or multiple issues occurred), or optimal quantification (without any limiting issues).

Definitions for limiting issues (Table 1) were discussed and agreed on by all authors. The investigators were then trained in the application of these criteria by scoring a separate data set that was discussed afterward. Feasibility scoring was then performed independently as described above. Disagreements were resolved in a consensus meeting, and the final score was used for further analysis. Interobserver agreement on the feasibility categories was calculated by using the intraclass correlation coefficient (SPSS Statistics 27, IBM), which was based on a mean-rating, absolute-agreement, two-way, mixed-effects model [16].

| Limiting issue | Description | # cases (%) |
|---|---|---|
| *Loss of correlation (systole)* | High velocities (and in some cases disturbed blood flow or high shear) during systole, causing low cross-correlation values during PIV analysis and subsequent loss of velocity vector accuracy. This issue was selected if areas with multiple vectors occurred with correlation values below 0.2 | 12/42 (29%) |
| *Short vessel segment* | Elongation of the blood vessels (or other anatomical reasons) causing part of the vessel to be outside the imaged plane. The vessel was scored as short, if the visible length was less than 4 diameters. | 8/42 (19%) |
| *Shadow regions* | Calcifications in the imaged atherosclerotic plaque causing darker regions (i.e. "shadows") in the ultrasound images. This issue was selected if contrast in the shadowed region was too low for adequate flow visualization. | 7/42 (16%) |
| *Loss of contrast (diastole)* | Destruction of the contrast microbubbles by the ultrasound, causing a severe loss of contrast in the diastolic phase. This issue was selected if all velocity vectors in the imaged region were lost at the end of diastole. | 5/42 (12%) |

**Table 1: Issues That Limited Flow Visualization. Data in parentheses are percentages. In some locations, multiple issues occurred; therefore, the total number is higher than the number of measured locations with partial flow visualization. PIV = particle image velocimetry.**

Temporal velocity profiles were acquired at five vector locations along the centerline of the vessel and were used to automatically select systolic and diastolic phases. Two feasibility parameters were then calculated for both phases. First, the contrast-to-back-ground ratio was obtained from the HFR-CEUS data by selecting multiple regions within and outside the imaged vessel (Fig 1) and calculating the ratio of the average contrast intensity in those regions. The average normalized cross-correlation value of the velocity vectors was used as a measure of confidence in the velocity vectors, with scores ranging from 0 to 1. Paired t tests were performed to review the difference between systole and diastole. $p < 0.05$ was considered indicative of statistically significant difference.

### *Patient & Lesion characteristics*

| Demographics | Median (range) |
|---|---|
| Age (years) | 67 (56-84) |
| BMI | 26.4 (20.9-38.5) |
| | # of cases |
| Sex (man/woman) | 21/13 |
| Rutherford category (grade 2/3)[10] | 23/12 |
| | |
| **Risk factors (SVS grading system[11])** | **Total (grade 1/2/3)** |
| Diabetes mellitus | 5 (2/3/0) |
| Smoking | 23 (5/13/5) |
| Hypertension | 15 (7/8/0) |
| Renal disease | 4 (0/3/1) |
| Hyperlipidaemia | 32 (0/0/32) |
| Cardiac disease | 5 (5/0/0) |
| Pulmonary disease | 1 (0/1/0) |
| | |
| **Location of the lesion*** | **# of cases** |
| Aortic bifurcation | 3 |
| Common iliac artery - Left | 7 |
| Common iliac artery - Right | 11 |
| Iliac bifurcation - Left | 2 |
| Iliac bifurcation - Right | 0 |
| External iliac artery - Left | 9 |
| External iliac artery - Right | 4 |
| | |
| **CTA** | **Median (range)** |
| vessel diameter before lesion (mm) | 7.2 (4.7-12.8) |
| vessel diameter after lesion (mm) | 7.8 (3.9-13.0) |
| Agatston calcium score[12] | 1196 (0-4560) |
| Cases with shadows in HFR-CEUS data | 1614 (107-4063) |
| Cases without shadows in HFR-CEUS data | 915 (0-4560) |

**Table 2. Demographic and clinical data of the 34 patients where HFR-CEUS data was acquired. BMI=body mass index, CTA=contrast-enhanced computed tomography. HFR-CEUS = high-frame-rate contrast-enhanced ultrasound. SVS = Society for Vascular Surgery.**
\* 36 lesions were measured in 34 patients, because 2 patients had a bilateral lesion.

*Flow Parameters*

Flow complexity, a measure of multidirectional blood flow [17], [18] and vorticity (ie, curl of the vectors [19]) were calculated in two participants with optimal blood flow quantification who showed a region with disturbed blood flow. Both flow parameters were compared with those from an undisturbed region in the same participant.

## Results

Thirty-five participants (22 men; median age, 67 years; age range, 56–84 years) were included (Fig 2, Table 2). HFR-CEUS measurements were obtained in 34 of the 35 participants (in one participant, the stenosis could not be visualized with use of either of the US machines). In eight participants, two separate locations were measured, resulting in 42 locations (Fig 2).

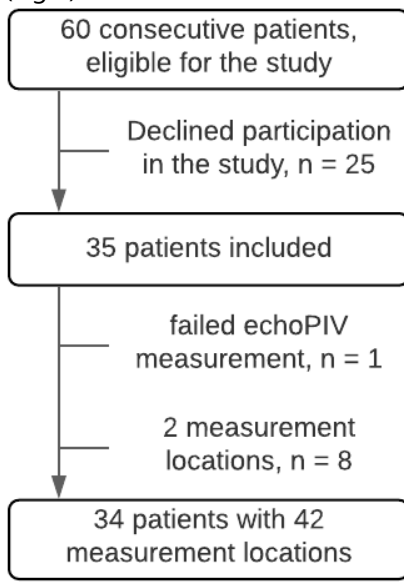

**Figure 2:** Flow diagram shows the inclusion of study participants and the number of measurement locations. echoPIV = US velocimetry.

### Feasibility

The intraclass correlation coefficient was 0.30 (95% CI: 0.21, 0.38). Flow quantification was achieved in 98% of measurements (41 of 42). In 38% (16 of 42 measurements), optimal flow quantification was achieved. In 60% (25 of 42) of measurements, only partial flow quantification was possible be-cause of loss of correlation during systole (n = 12) (Fig 3A, Movie 1 [online]), short vessel segments (n = 7) (Fig 3B), shadow regions (n = 8) (Figure 3C-D) and loss of contrast during diastole ($n=5$) (Figure 3E-F). Calcium scores in the locations where shadows occurred were higher than in other cases (Table 2).

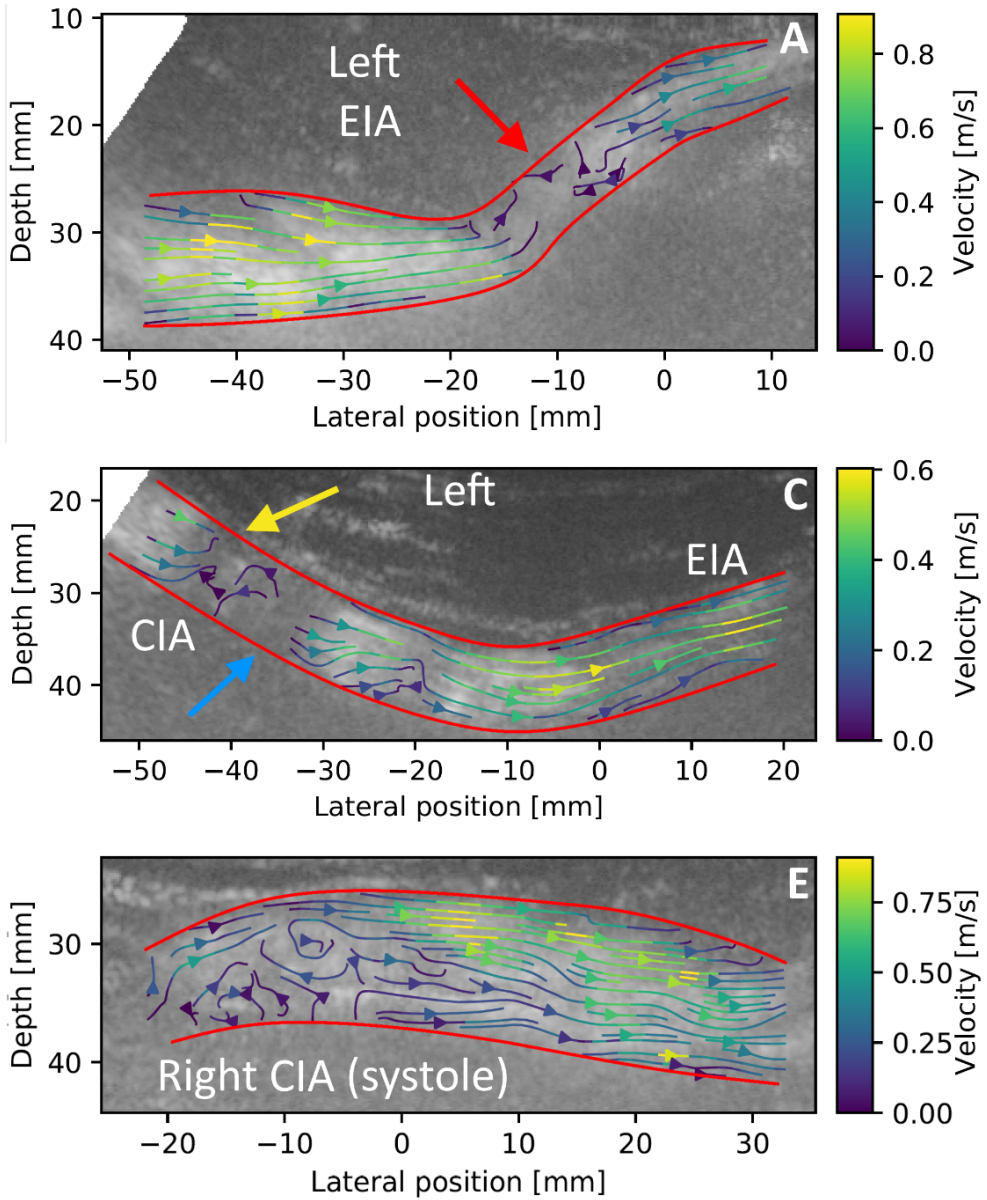

**Figure 3:** US velocimetry images show examples of "partial" flow visualization in four participants. Solid red lines indicate the borders of the vessel, and lines within these borders indicate the flow pattern (arrowheads indicate direction). All US images were obtained by placing the transducer on the patient's abdomen in the long axis of the corresponding vessel. (A) A stenotic lesion (red arrow) in the external iliac artery (EIA) of a 61-year-old man caused very fast and disturbed flow patterns. These patterns could not be adequately visualized. (B) The left common iliac vein (CIV) crosses the right

common iliac artery (CIA) in this 58-year-old man, leaving only short arterial vessel segments in the imaged plane.

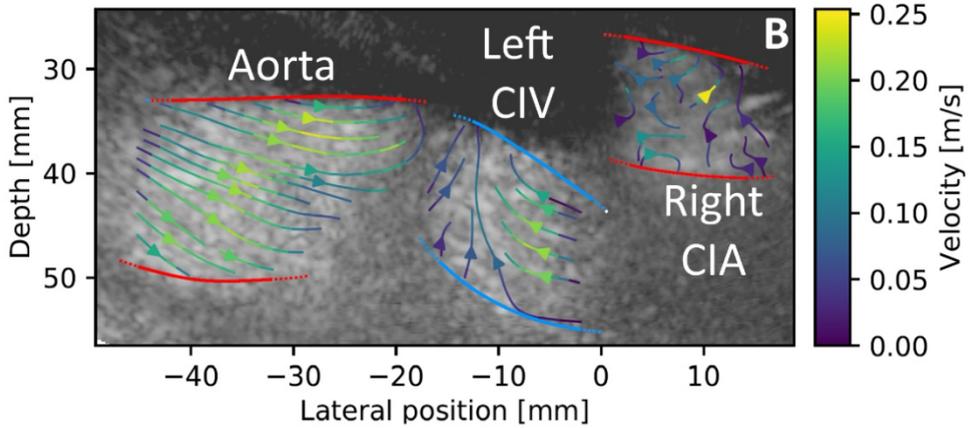

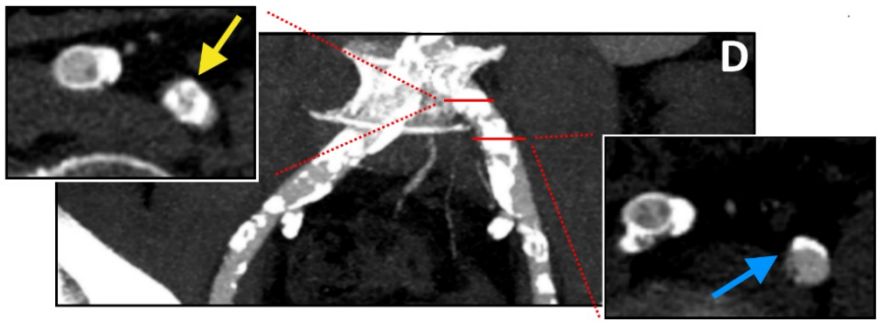

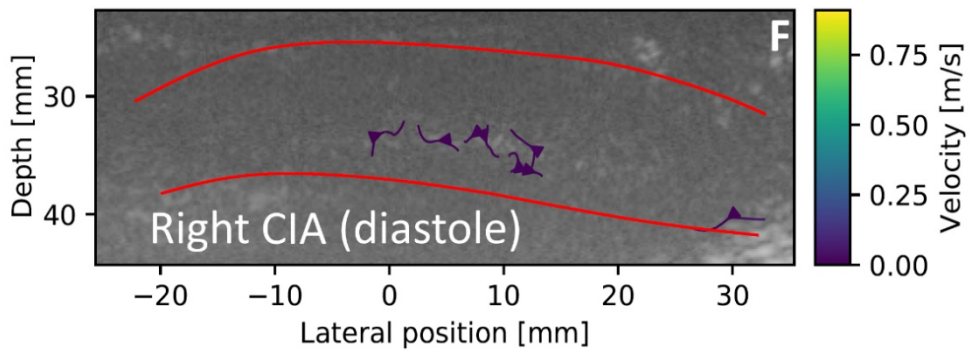

(C) Shadows caused by calcifications (yellow and blue arrows) limit contrast intensity, and consequently, the visualization of blood flow in this 69-year-old man. (D) Contrast-enhanced CT images of the same participant in C (coronal image with transverse sections at the location of the red lines) shows calcifications on the anterior side of the left common iliac artery (yellow and blue arrows). (E, F) Images in an 84-year-old woman with a stenosis in the proximal common iliac artery show poststenotic disturbed flow during systole (lateral position at around 210 mm). During diastole, the contrast microbubbles were destroyed, and flow visualization was not possible.

The contrast-to-background ratio was significantly higher during systole than diastole (11.0 6 2.9 vs 6.9 6 3.4, respectively; p < 0.001). The lowest contrast-to-background ratio values during diastole correspond to the five locations where loss of contrast was observed (Fig 4). The mean correlation values of the velocity vectors were significantly lower during systole (0.58 6 0.21 vs 0.47 6 0.13; p < 0.001), except in those same five locations.

## Flow Parameters

Flow complexity was higher in regions with disturbed flow (Figure 5A, green box). In one participant this difference was most pronounced during systole (Figure 5C, right side). Both participants showed a similar increase in vortical flow during systole in both regions, but vorticity was higher in the regions with disturbed flow.

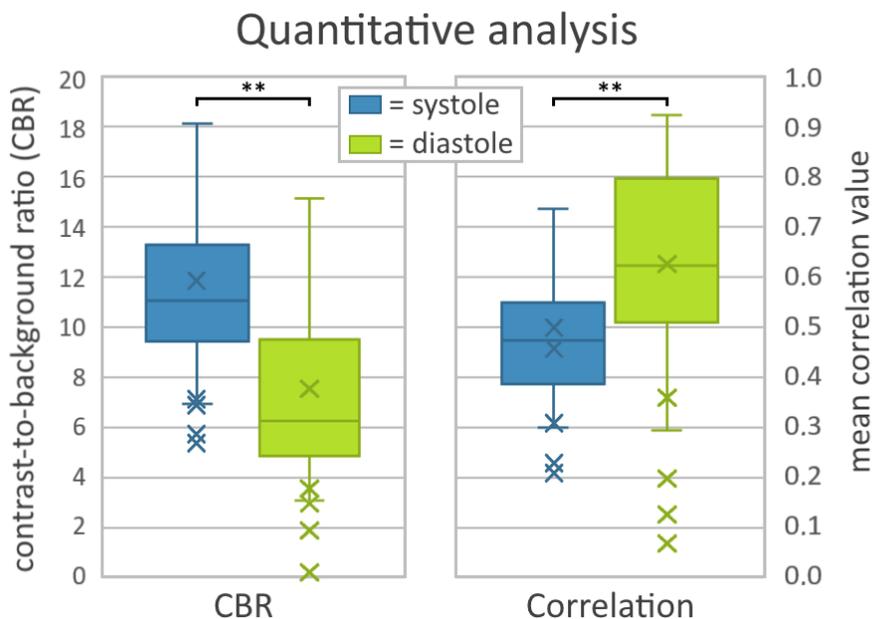

Figure 4: Contrast-to-background ratio (CBR) and mean vector correlation values during systole (blue) and diastole (green). Boxes indicate upper and lower quartiles, and whiskers indicate highest and lowest values. The five cases categorized as having severe contrast destruction at qualitative scoring are presented as separate markers but were included in the paired t-test. ** = For both parameters, there is a significant difference between systole and diastole. The mean contrast-to-background was lower during diastole (6.9 6 3.4 [standard deviation] vs 11.0 6 2.9; p <0.001), and the mean vector correlation value was higher during diastole (0.58 6 0.21 vs 0.47 6 0.13; p < 0.001).

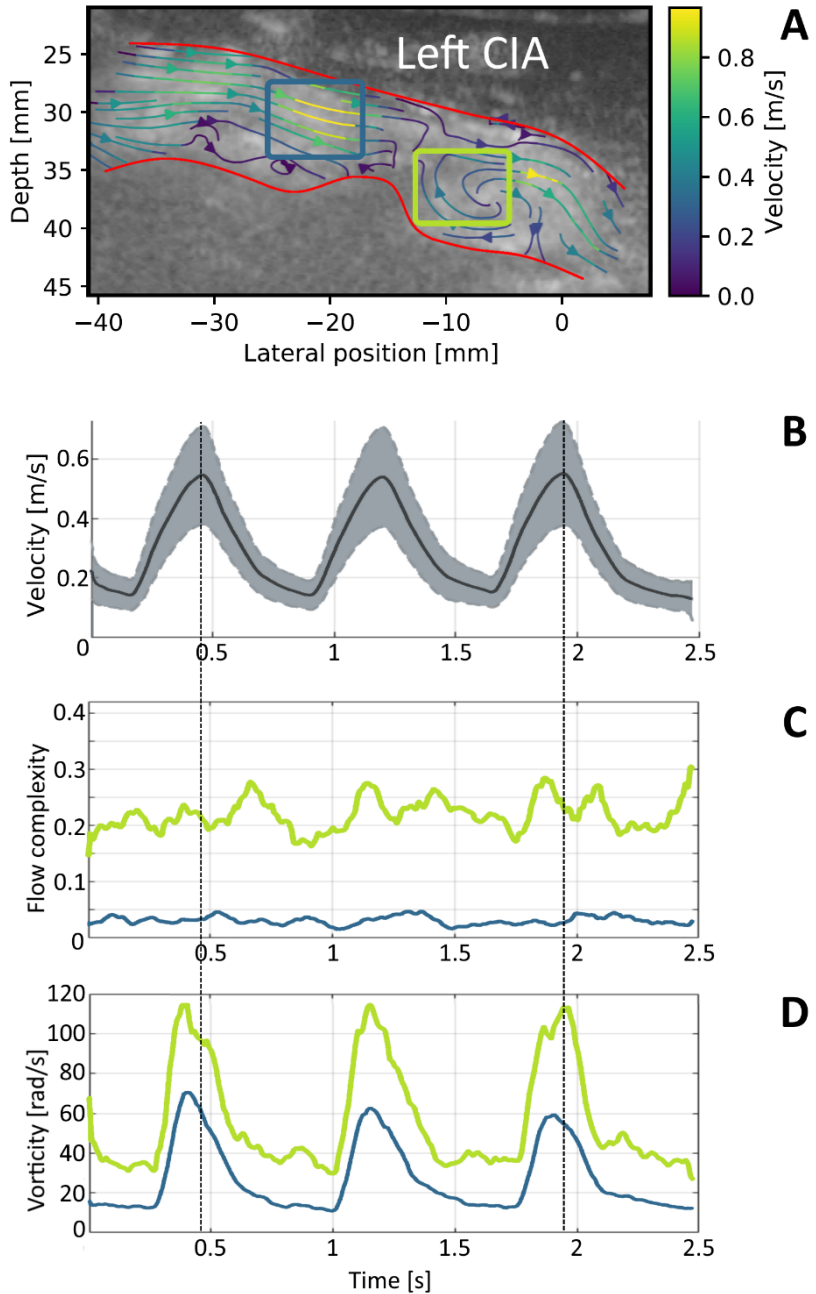

Figure 5: (A) US were images obtained by placing the transducer on the patient's abdomen and imaging the long axis of the corresponding vessel. US velocimetry data

were obtained during systole in two participants who had flow disturbances at visual inspection. Solid red lines indicate the borders of the vessel, and lines within these borders indicate the flow pattern (arrowheads indicate direction). Flow parameters were calculated in a region with undisturbed flow (blue boxes) and in a region with disturbed flow (green boxes).

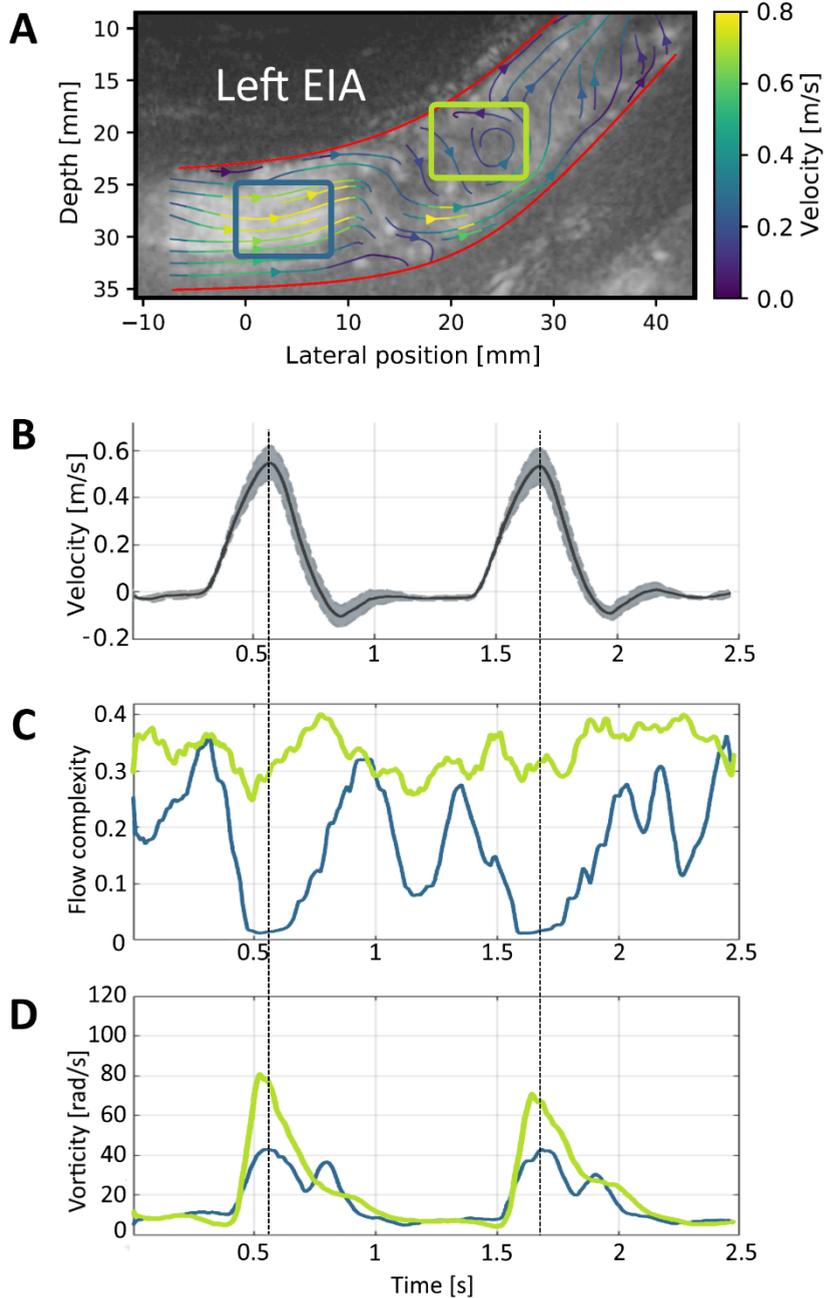

CIA = common iliac artery, EIA = external iliac artery. (B) Temporal velocity profiles acquired at five locations along the centerline of the vessel, showing several heart cycles. Shaded error bars represent the range of measured velocities. (C) Flow complexity and (D) vorticity in the region with undisturbed blood flow (blue) and in the region with disturbed blood flow (green). Left side of A–D: blood flow in the left common iliac artery in a 56-year-old woman. Right side of A–D: blood flow in the left external iliac artery in a 69-year-old man.

## Discussion

This study showed that US velocimetry, or echoPIV, in the aortoiliac tract was feasible in 98% of locations. Optimal quantification was achieved in 38% of locations, indicating that the technique needs to be optimized. In participants with optimal quantification, disturbed blood flow patterns could be clearly distinguished from undisturbed blood flow patterns in areas without an increased peak systolic velocity that were not identified as problematic areas at duplex US. Partial quantification was achieved in 60% of locations. Here, blood flow could still be visualized, but this was only possible during part of the cardiac cycle or in a subregion of the imaged vessel.

We previously showed a good match between echoPIV and phase-contrast MRI in the aortic bifurcation in healthy volunteers [9]. In the current study, echoPIV flow patterns closely matched the movement of the contrast microbubbles on high-frame-rate contrast-enhanced US (HFR-CEUS) images and are therefore assumed to be valid. In most participants, we were not able to obtain reliable Doppler velocity measurements because the maximum angle of 60° was not achieved or because the direction of the flow could not be accurately estimated. Therefore, these data could not be used as a reference. This also confirms the inherent limitations of Doppler imaging due to its angle dependency.

Without the use of duplex US, a reference standard to com-pare the blood flow velocities measured with echoPIV is lacking. A suitable alternative would have been to use phase-contrast MRI. However, this was not available at our institution during the study.

Despite thorough training, the interobserver agreement was still poor (Intraclass correlation coefficient, 0.30), emphasizing the complexity of this unvalidated quality assessment developed in-house. During the consensus meeting, disagreements were mostly caused by different interpretations of the exact cutoff for each limiting issue. Consensus was achieved more easily for each subsequent case, indicating that the interpretation of the scoring criteria converged. Rescoring would therefore likely improve interobserver agreement, but not the scoring method itself.

Optimal blood flow quantification was not achieved in all imaged vessel segments because of several limiting issues. Some of them, including calcifications and out-of-plane blood flow, affect US imaging in general. Loss of contrast (i.e., microbubble destruction) is a problem that is exacerbated with the use of HFR-CEUS because of the increased exposure of individual microbubbles to ultrasound waves. Severe bubble destruction occurred in five participants, whereas only minor destruction occurred in a previous study in healthy volunteers for whom similar US intensities were used [9]. This could be explained by stagnant blood flow during diastole in patients with atherosclerosis, which does not occur in healthy volunteers. Decreasing the mechanical index further to prevent this destruction would have resulted in an inadequate signal-to-noise ratio. This issue could be addressed by using novel contrast agents that are either more stable during insonification or produce stronger US reflections at a lower mechanical index [20].

In addition, echoPIV requires capturing an entire vessel segment in a single image plane. This is challenging in patients with atherosclerosis, who typically have elongated and curved arteries. Three-dimensional US acquisitions are needed to properly capture these out-of-plane vessels. However, this technique is still in early development [21], [22].

In our study, data processing and subsequent flow quantification were performed off-line, without having flow information as feedback to optimize measurement settings and transducer positioning. Translation of echoPIV to daily clinical practice would greatly benefit from real-time flow quantification capabilities, which would require direct data processing [23]. This could then be used to search for clinically relevant flow features, instead of relying on anatomic features.

Despite these technical limitations, flow disturbances were successfully quantified with use of echoPIV by measuring vector complexity and vorticity. These parameters could be used as an alternative for local wall shear stress values, which currently cannot be calculated accurately with use of echoPIV, and may predict atherosclerotic disease progression. In the future, longitudinal studies with a larger sample size will be needed to show the prognostic value and clinical impact of this technology.

# Conclusion

Blood flow quantification is feasible by using US velocimetry (echoPIV) in patients with aortoiliac stenosis. Technical challenges—such as microbubble stability, three-dimensional imaging methods, and direct data processing—must be addressed for clinical implementation. Nonetheless, echoPIV already enables acquisition of additional information, including vector complexity and vorticity, that can be used to distinguish disturbed from undisturbed blood flow in regions that are not identified as problematic areas by using duplex US.

## Acknowledgements

The authors thank Bastiaan Bongers, Laura Bouwmeister, Jochem Noordzij, Pinel Schrijver, and Frans Tak for their assistance during the high–frame-rate contrast-enhanced US measurements.

# References

[1] M. H. Friedman, G. M. Hutchins, C. Brent Bargeron, O. J. Deters, and F. F. Mark, "Correlation between intimal thickness and fluid shear in human arteries," *Atherosclerosis*, vol. 39, no. 3, pp. 425–436, 1981.

[2] D. N. Ku, D. P. Giddens, C. K. Zarins, and S. Glagov, "Pulsatile flow and atherosclerosis in the human carotid bifurcation. Positive correlation between plaque location and low and oscillating shear stress," *Arteriosclerosis*, vol. 5, no. 3, pp. 293–302, 1985.

[3] A. M. Malek, S. L. Alper, and Izumo, "Hemodynamic Shear Stress and Its Role in Atherosclerosis," *J. Am. Med. Assoc.*, vol. 282, no. 21, pp. 2035–2042, 1999.

[4] C. J. Slager *et al.*, "The role of shear stress in the generation of rupture-prone vulnerable plaques," *Nat. Clin. Pract. Cardiovasc. Med.*, vol. 2, no. 8, pp. 401–407, 2005.

[5] S. B. Coffi, D. T. Ubbink, I. Zwiers, A. J. M. Van Gurp, and D. A. Legemate, "The value of the peak systolic velocity ratio in the assessment of the haemodynamic significance of subcritical iliac artery stenoses," *Eur. J. Vasc. Endovasc. Surg.*, vol. 22, no. 5, pp. 424–428, 2001.

[6] S. G. Heinen *et al.*, "Hemodynamic significance assessment of equivocal iliac artery stenoses by comparing duplex ultrasonography with intra-arterial pressure measurementsitle," *J. Cardiovasc. Surg. (Torino).*, vol. 59, no. 1, pp. 37–44, 2018.

[7] R. W. Gill, "Measurement of blood flow by ultrasound: accuracy and sources of error," *Ultrasound Med. Biol.*, vol. I, no. 4, pp. 625–641, 1985.

[8] D. T. Ubbink, M. Fidler, and D. A. Legemate, "Interobserver variability in aortoiliac and femoropopliteal duplex scanning," *J. Vasc. Surg.*, vol. 33, no. 3, pp. 540–545, 2001.

[9] S. Engelhard *et al.*, "High-frame-rate contrast-enhanced US particle image velocimetry in the abdominal aorta: First human results," *Radiology*, vol. 289, no. 1, pp. 119–125, 2018.

[10] R. B. Rutherford *et al.*, "Recommended standards for reports dealing with lower extremity ischemia: Revised version," *J. Vasc. Surg.*, vol. 26, no. 3, pp. 517–538, 1997.

[11] M. C. Stoner *et al.*, "Reporting standards of the Society for Vascular Surgery for endovascular treatment of chronic lower extremity peripheral artery disease," *J. Vasc. Surg.*, vol. 64, no. 1, pp. e1–e21, 2016.

[12] R. Vaccarino, M. Abdulrasak, T. Resch, A. Edsfeldt, B. Sonesson, and N. V. Dias, "Low Iliofemoral Calcium Score May Predict Higher Survival after EVAR and FEVAR," *Ann. Vasc. Surg.*, vol. 68, no. April, pp. 283–291, 2020.

[13] G. Montaldo, M. Tanter, J. Bercoff, N. Benech, and M. Fink, "Coherent plane-wave compounding for very high frame rate ultrasonography and transient

elastography," *IEEE Trans. Ultrason. Ferroelectr. Freq. Control*, vol. 56, no. 3, pp. 489–506, 2009.

[14] C. Demene *et al.*, "Spatiotemporal Clutter Filtering of Ultrafast Ultrasound Data Highly Increases Doppler and fUltrasound Sensitivity," *IEEE Trans. Med. Imaging*, vol. 34, no. 11, pp. 2271–2285, Nov. 2015.

[15] J. Voorneveld *et al.*, "High Frame Rate Contrast-Enhanced Ultrasound for Velocimetry in the Human Abdominal Aorta," *IEEE Trans. Ultrason. Ferroelectr. Freq. Control*, vol. 65, no. 12, pp. 2245–2254, 2018.

[16] T. K. Koo and M. Y. Li, "A Guideline of Selecting and Reporting Intraclass Correlation Coefficients for Reliability Research," *J. Chiropr. Med.*, vol. 15, no. 2, pp. 155–163, 2016.

[17] A. E. C. M. Saris, H. H. G. Hansen, S. Fekkes, J. Menssen, M. M. Nillesen, and C. L. de Korte, "In Vivo Blood Velocity Vector Imaging Using Adaptive Velocity Compounding in the Carotid Artery Bifurcation," *Ultrasound Med. Biol.*, vol. 45, no. 7, pp. 1691–1707, 2019.

[18] M. M. Pedersen *et al.*, "Novel Flow Quantification of the Carotid Bulb and the Common Carotid Artery with Vector Flow Ultrasound," *Ultrasound Med. Biol.*, vol. 40, no. 11, pp. 2700–2706, 2014.

[19] J. Jensen *et al.*, "Accuracy and Precision of a Plane Wave Vector Flow Imaging Method in the Healthy Carotid Artery," *Ultrasound Med. Biol.*, vol. 44, no. 8, pp. 1727–1741, 2018.

[20] E. Stride *et al.*, "Microbubble Agents: New Directions," *Ultrasound Med. Biol.*, vol. 46, no. 6, pp. 1326–1343, 2020.

[21] J. Voorneveld *et al.*, "High-Frame-Rate Echo-Particle Image Velocimetry Can Measure the High-Velocity Diastolic Flow Patterns," *Circ. Cardiovasc. Imaging*, vol. 12, no. 4, p. e008856, 2019.

[22] J. Voorneveld *et al.*, "4-D Echo-Particle Image Velocimetry in a Left Ventricular Phantom," *Ultrasound Med. Biol.*, vol. 46, no. 3, pp. 805–817, 2020.

[23] E. Boni, A. C. H. Yu, S. Freear, J. A. Jensen, and P. Tortoli, "Ultrasound open platforms for next-generation imaging technique development," *IEEE Trans. Ultrason. Ferroelectr. Freq. Control*, vol. 65, no. 7, pp. 1078–1092, 2018.